\documentclass[12pt, draftclsnofoot, onecolumn]{IEEEtran}
\IEEEoverridecommandlockouts

\usepackage{algorithmic,amsmath,amssymb,amsthm,array,bbm,bibentry,cite,color,comment}
\usepackage{enumerate,enumitem,eurosym,float,graphicx,lettrine,mathrsfs,multirow,nomencl}
\usepackage{pict2e,psfrag,ragged2e,relsize,setspace,subfigure,supertabular,tabularx,url}
\usepackage[english]{babel}

\newtheorem{thm}{Theorem}

\newtheorem{remark}{Remark}

\newtheorem{cor}{Corollary}

\newcommand{\e}{\mathbf{e}}

\newcommand{\h}{\mathbf{h}}

\newcommand{\w}{\mathbf{w}}

\renewcommand{\H}{\mathbf{H}}
\newcommand{\I}{\mathbf{I}}

\newcommand{\setA}{\mathcal{A}}

\newcommand{\setK}{\mathcal{K}}

\newcommand{\setM}{\mathcal{M}}

\newcommand{\setR}{\mathcal{R}}
\newcommand{\setS}{\mathcal{S}}

\newcommand{\herm}{\mathrm{H}}

\newcommand{\snr}{\mathrm{snr}}

\newcommand{\ex}{\operatornamewithlimits{\mathsf{E}}}
\newcommand{\var}{\operatornamewithlimits{\mathsf{Var}}\!}
\newcommand{\bydef}{\stackrel{\cdot}{=}}

\renewcommand{\snr}{\mathsf{snr}}

\newcommand{\TAS}{\textnormal{\tiny{TAS}}}
\newcommand{\MRT}{\textnormal{\tiny{MRT}}}
\newcommand{\OSTBC}{\textnormal{\tiny{OSTBC}}}
\newcommand{\MIMO}{\textnormal{\tiny{MIMO}}}
\newcommand{\as}{\textnormal{\tiny{as}}}
\newcommand{\LSNR}{\textnormal{\tiny{L}}}
\newcommand{\HSNR}{\textnormal{\tiny{H}}}
\newcommand{\FB}{\textnormal{\tiny{FB}}}

\begin{document}
\title{Delay Performance of MISO Wireless Communications}

\author{Jes\'{u}s~Arnau,~\IEEEmembership{Member,~IEEE,}
	and~Marios~Kountouris,~\IEEEmembership{Senior Member,~IEEE}
	\thanks{The authors are with the Mathematical and Algorithmic Sciences Lab, France Research Center, Huawei Technologies France SASU, 20 Quai du Point du Jour, 92100 Boulogne-Billancourt, France.
 Email: (\{jesus.arnau,marios.kountouris\}@huawei.com).}
	}


\maketitle

\begin{abstract}
Ultra-reliable, low latency communications (URLLC) are currently attracting significant attention due to the emergence of mission-critical applications and device-centric communication. URLLC will entail a fundamental paradigm shift from throughput-oriented system design towards holistic designs for guaranteed and reliable end-to-end latency. A deep understanding of the delay performance of wireless networks is essential for efficient URLLC systems. In this paper, we investigate the network layer performance of multiple-input, single-output (MISO) systems under statistical delay constraints. We provide closed-form expressions for MISO diversity-oriented service process and derive probabilistic delay bounds using tools from stochastic network calculus. In particular, we analyze transmit beamforming with perfect and imperfect channel knowledge and compare it with orthogonal space-time codes and antenna selection. The effect of transmit power, number of antennas, and finite blocklength channel coding on the delay distribution is also investigated. Our higher layer performance results reveal key insights of MISO channels and provide useful guidelines for the design of ultra-reliable communication systems that can guarantee the stringent URLLC latency requirements.
\end{abstract}


\section{Introduction} \label{sec:intro}
Data traffic has been growing tremendously over the last decade, fueled by the ubiquity of smart mobile devices and bandwidth-demanding applications. 
In order to handle the ever-increasing traffic load, existing wireless networks have typically been designed and planned with a focus on improving spectral efficiency and increasing coverage. The latency requirements of different applications have mostly been an after-thought. 
Ultra-high reliability and low latency have not been in the mainstream in most wireless networks, due to the focus on human-centric communications, delay-tolerant content and reliability levels in the order of 95-99\%. However, a plethora of socially useful applications and new uses of wireless communication are currently envisioned in areas such as industrial control, smart cities, augmented and virtual reality (AR/VR), automated driving or flying, robotics, telemedicine, algorithmic trading and tactile Internet.
In response, new releases of mobile cellular networks (mainly 5G new radio and beyond) are envisaged to support ultra-reliable, low latency communications (URLLC) scenarios with strict requirements in terms of latency (ranging from 1 ms to few milliseconds end-to-end latency depending on the use cases) and reliability (higher than 99.9999\%). Another new feature is the support of machine-type communications (MTC), where a massive number of connected devices transmit reliably a relatively low volume of non-delay-sensitive payload.

Information theory and communication engineering have been instrumental in boosting spectral efficiency and approaching the capacity limits. Nevertheless, URLLC and device-centric communication pose significant theoretical and practical challenges, requiring a departure from capacity-oriented system design towards a holistic view (network architecture, control, and data) for guaranteed and reliable end-to-end latency. Applying information theory to the design of low latency networks has been a long-standing challenge \cite{Tony_union}. Information theory mostly focuses on asymptotic limits, which can be achieved with arbitrarily small probability of error using long codewords, hence arbitrarily large coding delays. Despite recent development on the block error rates for finite blocklength codes \cite{Polyanskiy10}, more work is needed to better understand the non-asymptotic fundamental tradeoffs between delay, reliability and throughput, including both coding delays and queueing delays. In addition, the highly variable and delay-sensitive nature of network traffic together with the associated overhead (metadata) should be incorporated in the conventional communication theoretic framework. 

Reliable communication is a well-studied problem, dating back to Shannon's landmark paper \cite{Shannon}, and diversity-achieving techniques are usually employed to increase reliability by combating or exploiting channel variations. Several schemes have been developed, including error correction codes, the use of multiple antennas, and space/frequency diversity at the physical layer, as well as automatic repeat-request (ARQ), opportunistic scheduling, and erasure coding at higher layers. Among them, multiple-input multiple-output (MIMO) systems have received great interest due to their potential to combat fading, increase spectral efficiency, and reduce interference. MIMO techniques can be used for beam steering, diversity, spatial multiplexing and interference cancellation. Diversity-achieving techniques increase reliability by combating or exploiting channel variations, while beam steering techniques increase received signal quality by focusing desired energy or attenuating undesired interference. Spatial multiplexing increases the data rate by transmitting independent data symbols across the antennas. In this work, we focus on diversity and beam steering techniques, as being more relevant for reliable communications. In particular, we consider maximum ratio transmission (MRT), a transmit beamforming technique that maximizes the received signal and realizes diversity exploiting channel state information (CSI) at the transmitter. 

There have been several attempts at addressing latency considerations in the physical layer, including error exponents \cite{ErrorExp}, delay-limited capacity \cite{DelayLimitCapa}, outage capacity \cite{Ozarow}, throughput-delay tradeoff curves \cite{TD_ElGamal}, and finite blocklength channel coding \cite{Polyanskiy10}. 
In networking, delay is a key performance measure and queueing theory has been instrumental in providing exact solutions for backlog and delays in packet-switched networks. However, queueing network analysis is largely restricted to single-queue networks, few interacting (coupled) queues, small topologies and Poisson arrivals. Classical queueing models typically allow the analysis of average delay, failing to characterize delay quantiles (worst-case delay) and distributions, which are of cardinal importance in mission-critical applications. Recent efforts to combine queueing with communication theory, such as stochastic network calculus \cite{Chang00,Jiang08book,Fidler15_Guide}, timely throughput \cite{Timely_Kumar}, effective bandwidth \cite{Chang00}, and effective capacity \cite{WuNegi03_EC} to name a few, take on a different approach and compute performance bounds for a wide range of stochastic processes. These approaches promise significant performance gains - in terms of latency, reliability and throughput - and crisp insights for the design of low latency communication systems. In this work, we employ stochastic network calculus - a probabilistic extension of (deterministic) network calculus \cite{Cruz91_TIT1} - which allows non-asymptotic stochastic bounds on network performance metrics, such as maximum delay, for broad classes of arrival, scheduling, and service processes. 

Despite the extended literature on MIMO techniques at the physical layer, only few attempts have been made to characterize the upper layer performance of multi-antenna techniques taking into account the queueing effects. In \cite{MIMO_AMC} the service process of an adaptive MIMO system with Poisson arrivals is characterized. Bounds on the delay violation probability have been derived for MIMO multiple access with bursty traffic in \cite{Javidi1}, while \cite{Javidi2} provides an asymptotic analysis of the diversity-multiplexing tradeoff for MIMO systems with bursty and delay-limited information. Using large deviations, \cite{ChenLau} analyzes the queueing performance of queue-aware scheduling in multiuser MIMO systems. Bounds on the tail of delay of MIMO communication systems have been derived using the effective capacity framework \cite{Gursoy,Matthaiou2012,MISO_Eff}. Nevertheless, these approximations are only valid for large delays and under constant bit rate processes. Using Markov chains to reproduce the state of Gilbert-Elliott fading channels, the flow-level performance of MIMO spatial multiplexing has been analyzed using stochastic network calculus in \cite{Mahmood1,Mahmood2}. Nevertheless, none of these works considered the delay performance of MIMO schemes using stochastic network calculus for wireless fading channels. 

In this work, we study the upper layer delay performance of multiple input, single output (MISO) diversity communication in the presence of statistical delay constraints. We consider MRT transmit beamforming at the physical layer and derive probabilistic delay bounds using tools from stochastic network calculus. For that, we provide a closed-form characterization of the cumulative service process for MISO beamforming channels with both perfect and imperfect CSI. For the analysis, we use the (min,$\times$) network calculus methodology for fine-grained wireless network delay analysis \cite{Multihop13_Infocom}. The impact of transmit antennas, signal-to-noise ratio (SNR), and imperfect CSI on the delay distribution of MISO MRT systems is characterized. We then show that our mathematical framework can be applied to the statistical characterization of various MIMO service processes, including MIMO eigen-beamforming, orthogonal space-time block coding (OSTBC), antenna selection, and Nakagami-$m$ fading channels. This allows us to compare the delay performance of transmit beamforming with alternative diversity-achieving techniques that rely on very low rate CSI (transmit antenna selection) or no CSI (OSTBC). Interestingly, MISO MRT is shown to reduce the delay violation probability as compared to single-antenna transmissions even with imperfect CSI. The derived delay bounds enable us to assess the robustness of MISO MRT delay performance with respect to channel imperfections. Our results also show under which operating parameters other diversity-techniques are preferable than MRT in terms of delay violation probability. In addition, we provide an asymptotic statistical characterization of the service process in the low/high SNR regime and for large number of antennas. Finally, extending \cite{Al-Zubaidy15_FiniteBlock} to MISO systems, we study the effect of finite blocklength channel coding on the queueing delay performance. Our results quantify the performance loss due to finite blocklength and characterize the tradeoff between data rate and the error probability with respect to the delay performance.
Our results can provide useful insights and guidelines for the design of ultra-reliable wireless systems that can satisfy and guarantee the stringent URLLC latency requirements.

The rest of the paper is organized as follows: In Section \ref{sec:sys_model}, we provide our system model and in Section \ref{sec:SNC}, a brief background on the (min,$\times$) network calculus is presented. In Section \ref{sec:perf_shannon}, the delay performance analysis of MISO diversity systems is derived. Section \ref{sec:perf_shannon_asym} provides the delay performance in asymptotic regimes and Section \ref{sec:perf_FB} shows the effect of finite blocklength on the delay performance. Numerical results are presented in Section \ref{sec:num_results}, followed by conclusions in Section \ref{sec:conclusions}.

\section{System model} \label{sec:sys_model}
We consider data transmission over a point-to-point vector communication channel. Time is divided into time slots of duration $T$ (discrete-time model), and at each slot $i$, the source generates $a_i$ data bits and stores them in a queue.
The transmitter (source) has $M$ transmit antennas and sends the queued data bits to a single-antenna receiver over a frequency-flat Rayleigh fading channel. We assume a block-fading model, where the channel remains constant during one slot and varies independently from slot to slot. Each slot contains $n+n_m$ symbols, where $n$ denotes the complex data symbols and $n_m$ the metadata (headers, training, estimation, acknowledgments, etc.). 

\subsection{Signal model}
The received downlink signal $y_i \in \mathbb{C}$ at slot $i$ in a MISO wireless channel is given by
\begin{eqnarray}
y_i = \sqrt{\snr}\cdot \mathbf{h}_i^{\rm H}\mathbf{x}_i + n_i
\end{eqnarray}
where $\mathbf{h}_i \in \mathbb{C}^{M\times1}$ is the flat-fading channel between the transmitter and the receiver at the $i$-th slot, which is circularly-symmetric complex Gaussian distributed $\mathbf{h} \sim \mathcal{CN}(0,1)$. The transmitted vector is denoted by $\mathbf{x}_i \in \mathbb{C}^{M\times1}$, and $n_i \sim \mathcal{CN}(0,1)$ is the additive background noise that may also include (Gaussian) interference from neighboring systems. 
We consider one of the most prominent multi-antenna diversity technique, namely transmit beamforming, which refers to sending linearly weighted versions of the same signal on each antenna. The transmitted signal can be written as $\mathbf{x}_i = \mathbf{w}_is_i$, where $s_i$ is the zero-mean data signal at slot $i$ with power $\ex\left[|s|^2\right] = 1$, and $\mathbf{w}_i \in \mathbb{C}^{M\times1}$ is the unit-norm beamforming vector. Note that, since noise is assumed to have unit power, $\snr$ represents the average received SNR, whereas the instantaneous SNR in the $i$-th slot is given by $\gamma_i = \snr|\mathbf{h}_i^{\rm H}\mathbf{w}_i|^2$. 

\subsection{Transmission mode}\label{subsec:tx_mode}
A natural signaling strategy for the MISO channel will be to maximize SNR over a specific channel, which can be achieved by sending information only in the direction of the channel vector $\mathbf{h}$, as information sent in any orthogonal direction will be nulled out by the channel anyway. We thus consider the so-called maximum ratio transmission (MRT) \cite{Lo1999}, which is equivalent to eigen-beamforming since beamforming along the dominant (and only) eigenmode of the $M \times 1$ vector channel is performed\footnote{It can be shown that MRT is the transmission scheme that maximizes both the capacity and the service rate with sum power constraint, Gaussian input signaling, and perfect CSI. This is not necessarily true for arbitrary inputs, where the transmission scheme may depend on the delay constraints.}. 

Assuming that both transmitter and receiver have perfect CSI, the MRT beamforming vector is given by $\mathbf{w}_i = \frac{\mathbf{h}_i}{\left\|\mathbf{h}_i\right\|}$. In that case, the instantaneous SNR is $\gamma_i = \snr\|\mathbf{h}_i\|^2$, which is gamma distributed with shape parameter $M$ and scale parameter $\snr$, i.e. $\gamma_i \sim \mathrm{Gamma}(M,\snr)$.
When the transmitter does not fully know the actual channel vector $\h$ (imperfect CSI), we can model its channel knowledge as $\hat{\h} = \h+\e$, where $\e\sim\mathcal{CN}(0,\sigma_\mathrm{e}^2\I)$. MRT is then performed based on the channel estimate, so that $\w = \frac{{\hat{\h}}}{{\left\|\hat{\h}\right\|}}$. Particularizing \cite[Eq. 7]{Chen2005} to the MISO case, the instantaneous SNR is gamma distributed with shape parameter $M$ and scale parameter $\zeta$, i.e.
\begin{equation}\label{eq:snr_is_gamma}
\gamma_i\sim\mathrm{Gamma}(M,\zeta) \quad \text{with} \quad \zeta = \left(\sigma_\mathrm{e}^2+\frac{1+\sigma_\mathrm{e}^2}{\snr}\right)^{-1}.
\end{equation}
This additive error model is consistent with time-divison duplex (TDD) operation, where uplink and downlink transmissions take place at the same frequency, in different time instants; assuming they fall within the coherence interval of the channel, then channel reciprocity can be used to estimate the downlink channel from uplink pilot signals. This model also applies to frequency-division duplex (FDD) operation with analog feedback \cite{Caire2010}. We only account for the effect of CSI error in MISO beamforming, which reduces the achieved SNR (SNR loss) because of not transmitting exactly in the direction of the actual channel; as we explain in the next subsection there could be another penalty in the rate selection process.

\subsection{Data transmission}
A codeword of length $n$ symbols (corresponding to $n$ channel uses) and rate $R_i$ (in bits per symbol) is transmitted at each slot $i$. The transmitter selects a rate adapted to $\gamma_i$ and we consider the following two cases:

\subsubsection{Asymptotically large blocklength}
If the blocklength is large enough, no errors occur and the achievable rate is equal to the Shannon capacity of the channel, $R_i = \log_2(1+\gamma_i)$.

\subsubsection{Finite blocklength}
At finite blocklength, a transmission error can occur with probability $\epsilon > 0$ and the maximum coding rate $R_i(n, \epsilon)$ is lower than the Shannon rate. Tight non asymptotic upper and lower bounds on the maximum coding rate are given in \cite{Polyanskiy10}. Furthermore, for AWGN channels an asymptotic approximation has been established and shown accurate for packet sizes as small as 100 \cite{Polyanskiy10,Yang14_QuasiStatic,Hayashi09}. The coding rate $R_i(n, \epsilon) = k/n$ to transmit $k$ information bits using coded packets spanning $n$ channel uses is given by
\begin{eqnarray}
\label{eqn:RateFB}
R_i(n, \epsilon) \approx \log_2(1+\gamma_i) - \sqrt{\frac{V(\gamma_i)}{n}}Q^{-1}(\epsilon) + \frac{\log_2n}{2n}+ \mathcal{O}(1)
\end{eqnarray}     
where $Q^{-1}(\cdot)$ is the inverse of the Gaussian $Q$ function and $V(\gamma_i)$ is the channel dispersion given by
\begin{eqnarray}
V(\gamma_i) = \frac{\gamma_i(\gamma_i+2)}{(1+\gamma_i)^2}\log_2^2e.
\end{eqnarray} 

Using this approximation, the coding rate and the packet error probability are related as
\begin{eqnarray}
\epsilon = Q\left(\frac{n\log_2(1+\gamma_i) - k+0.5\log_2n}{\sqrt{V(\gamma_i)n}}\right). 
\end{eqnarray} 

The above approximation is valid for MISO systems with knowledge of the fading coefficients of the vector channel and the SNR realization at each slot, which makes the MISO channel behaving equivalently to an AWGN channel with SNR $\snr\|\h\|^2$. We should remark that so far throughout this section we have assumed perfect knowledge of the SNR realization, so that the transmitter can adapt the rate to it with no errors. Thus, we only account for the channel estimation error as an SNR penalty, as described in Sec.~\ref{subsec:tx_mode}. The case of imperfect rate selection in MISO systems, which goes beyond the scope of this work, can be analyzed using techniques recently developed in \cite{Schiessl2016}. In that case, channel estimation errors will be approximated by Gaussian variations in the SNR and the Gaussian variations in the capacity due to finite blocklength in \cite{Polyanskiy10} will be transformed into Gaussian errors in the SNR.

\subsection{Queuing model}\label{sec:queue_model}
For the analysis of queuing systems, we consider a stochastic system-theoretic model as in \cite{Multihop13_Infocom}, which is widely used in the stochastic network calculus methodology. Stochastic network calculus considers queuing systems and networks of systems with stochastic arrival, departure, and service processes, as the ones below. For an in-depth description of the topic, the interested reader may refer to \cite{Jiang05,Jiang08book,Fidler06_e2e, Fidler06GC,Fidler15_Guide,Multihop13_Infocom}.

The arrival process $a_i$, introduced in Sec.~\ref{sec:sys_model}, models the number of bits that arrive at the queue at a discrete time instant $i$. For successful transmissions, the service process $s_i$ is equal to: (i) $n R_i$ for asymptotically large blocklength, (ii) $n R_i(n, \epsilon)$ with finite blocklength; in case of transmission errors, the service is considered to be zero as no data is removed from the queue. Finite blocklength channel coding affects the reliability of the physical layer, which in turn causes additional delay as data needs to be buffered until successfully transmitted. 
The departure process $d_i$ describes the number of bits that arrive successfully at the destination and depends on both the service process and the number of bits waiting in the queue. Note that acknowledgments and feedback messages are assumed to be instantaneous and error-free.  

We further define the cumulative arrival, service and departure processes as
\begin{eqnarray}
A(\tau, t) = \displaystyle \sum_{i = \tau}^{t-1}a_i, \ \ S(\tau, t) = \displaystyle \sum_{i = \tau}^{t-1}s_i, \ \ D(\tau, t) = \displaystyle \sum_{i = \tau}^{t-1}d_i.
\end{eqnarray} 

For lossless first-in first-out queuing systems, the delay $W(t)$ at time $t$, i.e. the number of slots it takes for an information bit arriving at time $t$ to be received at the destination, is defined as 
\begin{eqnarray}
W(t) = \inf\{u > 0 : A(0,t)/D(0,t+u) \leq 1 \}.
\end{eqnarray} 
and the delay violation probability is given by $\Lambda(w,t) = \mathbb{P}\left[W(t) > w \right]$.

Using the dynamic server property (i.e. $D(0,t) \geq A*S(0,t)$ where the $(\min,+)$ convolution operator '*' is defined as $f*g(\tau,t) = \inf_{\tau \leq u \leq t}{f(\tau,t)+g(u,t)}$ \cite{Chang00}), the delay can be characterized through the cumulative arrival and service processes, which we have so far described in the so-called bit domain. As it is more convenient for the analysis of wireless fading channels, we follow \cite{Multihop13_Infocom} and analyze these processes in the exponential (or SNR) domain.

\section{Stochastic Network Calculus in the SNR Domain}\label{sec:SNC}

A remarkable feature of stochastic network calculus in the SNR domain is that it allows to obtain bounds on the delay violation probability based on simple statistical characterizations of the arrival and service processes in terms of their Mellin transforms. We will briefly review this result in this section.

Let us start by converting the cumulative processes in the bit domain through the exponential function. The corresponding processes in the SNR domain, denoted by calligraphic letters, are
\begin{equation}
\mathcal{A}(\tau, t) = e^{A(\tau, t)}, \quad \mathcal{D}(\tau, t) = e^{D(\tau, t)}, \quad \mathcal{S}(\tau, t) = e^{S(\tau, t)}.
\end{equation}

From these definitions, an upper bound on the delay violation probability can be computed by means of the Mellin transforms of $\mathcal{A}(\tau, t)$ and $\mathcal{S}(\tau, t)$:
\begin{equation}\label{eq:delay_bound_1}
  p_\mathrm{v}(w)  = \inf_{s>0}\left\lbrace K(s,-w)\right\rbrace \geq \Lambda(w)
\end{equation}
where $K(s,-w)$ is the so-called steady-state kernel, defined as
\begin{equation}\label{eq:ker_limits}
\mathcal{K}(s,-w) = \lim_{t\to\infty} \sum_{u=0}^{t}\mathcal{M}_{\mathcal{A}}(1+s,u,t)\mathcal{M}_{\mathcal{S}}(1-s,u,t+w).
\end{equation}
where $\mathcal{M}_X(s) = \ex\left[X^{s-1}\right]$ denotes the Mellin transform of a nonnegative random variable $X$ for any $s \in \mathbb{C}$ for which the expectation exists. We restrict our derivations in this work to $s \in \mathbb{R}$ and we recall that, for a continuous probability density function (pdf) $f_X(x)$ on $(0,\infty)$, if there exists $\delta > 0$ such that $\displaystyle \lim_{x\to0^+}\frac{f_X(x)}{x^{\delta+\theta-1}} < \infty$, then $\ex[X^{-\theta}]<\infty$. Alternatively, $\ex[X^{-\theta}]<\infty$ if and only if $\displaystyle \int_{0}^{\infty}\frac{F_X(x)}{x^{\theta+1}}{\rm d}x < \infty$. 

\subsection{Mellin transform of arrival and service processes} 
Assuming that $\mathcal{A}(\tau, t)$ has stationary and independent increments, the Mellin transforms become independent of the time instance, as follows:
\begin{eqnarray}
\setM_\setA (s,\tau,t) &=& \ex\left[\left(\prod_{i=\tau}^{t-1} e^{a_i}\right)^{s-1}\right] \\
&=& \ex\left[e^{a(s-1)}\right]^{t-\tau} \\
&=& \setM_\alpha(s)^{t-\tau}
\end{eqnarray}
where we have defined $\alpha = e^a$, the non-cumulative arrival process in the SNR domain.
We consider the traffic class of $(z(s), \rho(s))$-bounded arrivals, whose moment generating function in the bit domain is bounded by \cite{Chang00}
\begin{eqnarray}
\frac{1}{s}\log\ex[e^{sA(\tau,t)}] \leq \rho(s)\cdot(t-\tau) + z(s) 
\end{eqnarray}
for some $s>0$. Restricting ourselves to the case where $\rho$ is independent of $s$ and $z(s) = 0$, we have \cite{Al-Zubaidy15_FiniteBlock,Petreska2014}
\begin{equation}\label{eq:defmellin_alpha}
\setM_\alpha(s) = e^{\rho(s-1)}.
\end{equation}

For the service process, we start by rewriting $s_i = B\log g(\gamma)$, where $B = n/\log 2$. Since the different $s_i$ are independent and identically distributed (i.i.d.), we can express the Mellin transform of the cumulative service as
\begin{eqnarray}\label{eq:defmellin_s}
\setM_\setS (s,\tau,t) &=& \ex\left[\left(\prod_{i=\tau}^{t-1}g(\gamma)^B\right)^{s-1}\right] \\
					&=& \ex\left[g(\gamma)^{B(s-1)}\right]^{t-\tau} \\
					&=& \setM_{g(\gamma)}\left(1+B(s-1)\right)^{t-\tau}.
\end{eqnarray}

\subsection{Delay Bound}\label{sec:delay_viol}

Plugging (\ref{eq:defmellin_alpha}) and (\ref{eq:defmellin_s}) into (\ref{eq:ker_limits}) and following \cite{Multihop13_Infocom}, the steady-state kernel can be finally rewritten as 
\begin{eqnarray}
\mathcal{K}(s,-w) = \frac{\mathcal{M}_{g(\gamma)}(1-B\cdot s)^{w}}{1 - \mathcal{M}_{\alpha}(1+s)\mathcal{M}_{g(\gamma)}(1-B\cdot s)},
\end{eqnarray} 
for any $s > 0$ under the stability condition $\mathcal{M}_{\alpha}(1+s)\mathcal{M}_{\mathcal{S}}(1-s) < 1$. The delay bound (\ref{eq:delay_bound_1}) thus reduces to
\begin{equation}
p_\mathrm{v}(w) = \inf_{s>0}\left\lbrace\frac{\mathcal{M}_{g(\gamma)}(1-B\cdot s)^{w}}{1 - \mathcal{M}_{\alpha}(1+s)\mathcal{M}_{g(\gamma)}(1-B\cdot s)}\right\rbrace.
\end{equation}

\section{Delay with Large Blocklength: Exact Analysis} \label{sec:perf_shannon}
In this section, we derive exact closed-form expressions for the steady-state kernel $\setK(s,-w)$ of MISO diversity schemes when the blocklength is infinitely large. We start by providing a general result on the Mellin transform of the service process when the instantaneous SNR is gamma distributed. Obtaining the steady-state kernel for MRT beamforming with both perfect and imperfect CSI is a particularization of this result, which is shown to apply, as a byproduct, for obtaining the performance of other diversity techniques, including MISO OSTBC, antenna selection, and MIMO MRT/MRC. 

Consider the instantaneous SNR to be a gamma distributed random variable $\gamma \sim {\rm Gamma}(M,\zeta)$ with shape parameter $M$, scale parameter $\zeta$ and pdf
\begin{eqnarray}
f_{\gamma}(x) = \frac{x^{M-1}e^{-\frac{x}{\zeta}}}{\Gamma(M)\zeta^M}, \quad x \geq 0
\end{eqnarray} 
where $\Gamma(t) = \int_{0}^{\infty}x^{t-1}e^{-x}\,\mathrm{d}x$ is the (complete) gamma function; we have dropped the subindex since SNRs are independent and ergodic. First, we derive the Mellin transform of $g(\gamma)$, i.e. $\mathcal{M}_{g(\gamma)}(s) = \ex\left[g(\gamma)^{s-1}\right]$. For notation convenience, in the remainder we assume $B=n/\log2=1$, however in Sec.~\ref{sec:num_results} we give again relevant values to this parameter in order to obtain meaningful numerical results.

\begin{thm}\label{Th1:main}
The Mellin transform of $g(\gamma) = 1+\gamma$, where $\gamma \sim {\rm Gamma}(M,\zeta)$ with $M \in \mathbb{N}^+$ and $\zeta > 0$, is given by
\begin{eqnarray}\label{eq:mellin_miso_tricomi}
\mathcal{M}_{g(\gamma)}(s) = \zeta^{-M}\cdot U(M,M+s,\zeta^{-1})
\end{eqnarray}
where $U(a,b,z)$ is Tricomi's confluent hypergeometric function \cite[Eq. 13.2.5]{Abramowitz1964} (also called confluent hypergeometric function of the second kind and denoted by $\Psi(a;b;z)$).
\end{thm}
\begin{IEEEproof}
	See Appendix~\ref{sec:app_Th1main}.
\end{IEEEproof} 

\subsection{MISO MRT}

The Mellin transform derived above applies directly to the service process with MISO MRT transmission. Using this expression together with the transform of the arrival process, we obtain the kernel and consequently the bound on the delay violation probability as follows
\begin{eqnarray}\label{eq:pv_tricomi}
p_\mathrm{v}(w) & = & \inf_{s>0}\left\lbrace\frac{\mathcal{M}_{g(\gamma)}^\MRT(1-B\cdot s)^{w}}{1 - \mathcal{M}_{\alpha}(1+s)\mathcal{M}^\MRT_{g(\gamma)}(1-B\cdot s)}\right\rbrace \nonumber \\
& = & \inf_{s>0}\left\lbrace\frac{\left(\zeta^{-M}\cdot U(M,M+1-s,\zeta^{-1})\right)^{w}}{1-e^{\rho s}\zeta^{-M}\cdot U(M,M+1-s,\zeta^{-1})}\right\rbrace.
\end{eqnarray}

Although $U(a,b,z)$ is implemented in standard software for mathematical calculations, we provide below an alternative expression for the Mellin transform in terms of the simpler upper incomplete gamma function. 

\begin{thm}\label{Th2:main}
	The Mellin transform of $g(\gamma)$ from Theorem~\ref{Th1:main} can be given as
	\begin{eqnarray}\label{eq:mellin_miso_sum}
	\mathcal{M}_{g(\gamma)}(s)&=& \frac{e^\frac{1}{\zeta}}{\zeta^M\Gamma(M)}\sum_{j=0}^{M-1}\binom{M-1}{j}(-1)^{M-1-j}\cdot\zeta^{j+s}\cdot\Gamma(j+s,\zeta^{-1}) \\ 
	&=& e^\frac{1}{\zeta}\sum_{j=0}^{M-1}(-1)^{M-1-j}\cdot\zeta^{j+s-M}\frac{\Gamma(j+s,\zeta^{-1})}{\Gamma(M-j)\Gamma(j+1)}
	\end{eqnarray}
	where $\Gamma(s,z) = \int_{z}^{\infty}t^{s-1}e^{-t}\,\mathrm{d}t$ is the upper incomplete gamma function.
\end{thm}
\begin{IEEEproof}
	See Appendix~\ref{sec:app_Th2main}.
\end{IEEEproof}

\begin{remark}\rm
	For the SISO case, letting $M=1$ and $\zeta=\snr$ in (\ref{eq:mellin_miso_sum}) we obtain $\mathcal{M}_{g(\gamma)}(s) = e^{\frac{1}{\snr}}\cdot \snr^{s-1}\cdot\Gamma(s,\snr^{-1})$ which is the same expression reported in \cite{Multihop13_Infocom}. 
\end{remark}

The above expressions allows us to obtain bounds on the delay violation probability for different system parameters without resorting to Monte Carlo simulations. However, due to the complexity of the kernel function, no closed-form solution for the minimum $s$ can be found, and we must resort to numerical methods. In some asymptotic cases, we can have simpler expressions of the Mellin transform that make this process easier, as we will show later in Section~\ref{sec:perf_shannon_asym}.
 
\subsection{Other MISO Diversity Techniques}\label{sec:delay_diversity}
So far, we have considered that the transmitter performs MRT based on perfect or imperfect CSI. In this section, and for means of comparison, we study two alternative multi-antenna diversity techniques, namely OSTBC and transmit antenna selection, which rely on no and very low-resolution CSI, respectively. 

\subsubsection{OSTBC}\label{sec:delay_OSTBC}
Orthogonal space-time block coding has been a very successful transmit diversity technique because it can achieve full diversity without CSI at the transmitter and need for joint decoding of multiple symbols. It is characterized by the number of independent symbols $N_s$ transmitted over $T$ time slots; the code rate is $R_c = N_s/T$. When the transmitter uses OSTBC with $M$ transmit antennas, code parameter $T$, and the receiver performs MRC with $N$ antennas, the equivalent SNR $\gamma = \frac{\snr}{M}\|\mathbf{H}\|^2_{\rm F}$ is gamma distributed with shape parameter $MN$ and scale parameter $(\snr/M)^{-1}$ \cite[Eq. 3.43]{Paulraj2003}; here $\H$ denotes the MIMO channel matrix of $N\times M$ complex Gaussian entries. Particularizing (\ref{eq:mellin_miso_tricomi}) for the case of MISO OSTBC, we have the following result.
\begin{cor}\label{cor:OSTBC}
The Mellin transform of the service process of a MISO system employing OSTBC is given by
\begin{equation}
\setM_{g(\gamma)}^\OSTBC(s) = \left(\frac{\snr}{M}\right)^{-M}\cdot U\left(M,M+s,M/\snr\right).
\end{equation}
\end{cor}

\subsubsection{Antenna Selection}\label{sec:delay_OSTBC}
Antenna selection is a low-complexity, low-rate feedback diversity technique, in which the transmitter and/or the receiver select a subset of transmit/receive antennas for transmission/reception. It can be used in conjunction with other diversity techniques and can improve the performance of open-loop MIMO at the expense of very low amount of feedback. We consider here transmit antenna selection (TAS), in which the transmitter selects to transmit on the antenna (one of $M$) that maximizes the instantaneous SNR.
The amount of CSI required to be fed back to the transmitter is $\left \lceil{\log_2 M}\right \rceil$ bits (index of best antenna), where $\left \lceil{x}\right \rceil$ denotes the smallest integer larger than $x$. The instantaneous SNR can be expressed as $\gamma_{\rm TAS} = \snr\gamma_{\rm max}$, where $\gamma_{\rm max}$ is the largest channel gain, i.e. $\gamma_{\rm max} = \displaystyle \max_{1\leq i\leq M}|h_i|^2$.
Since $h_i\sim\mathcal{CN}(0,1)$, we have that $|h_i|^2$ is exponentially distributed with unit mean and pdf $f_{|h_i|^2}(x) = e^{-x}$.

\begin{thm}\label{Th3:main}
	The Mellin transform for a MISO system employing TAS is given by
	\begin{eqnarray}\label{eq:mellin_TAS}
	\mathcal{M}_{g(\gamma)}^\TAS(s) = M\zeta^{s-1}\sum_{k=0}^{M-1}\binom{M-1}{k}(-1)^{k}\frac{e^{\frac{k+1}{\zeta}}}{(k+1)^s}\Gamma\left(s,\frac{k+1}{\zeta}\right).
	\end{eqnarray}
\end{thm}
\begin{IEEEproof}
	See Appendix~\ref{sec:app_Th3main}.
\end{IEEEproof}

\subsection{Other applications of Theorem 1}\label{sec:delay_MIMO}
In this section we briefly point out towards other possible applications of Theorem 1.

\subsubsection{SISO case with Nakagami-$m$ fading}
	Theorem~\ref{Th1:main} could easily be used to analyze the SISO case with Nakagami-$m$ fading. The Nakagami-$m$ distribution includes as special cases Rayleigh ($m$ = 1), no fading ($m \to \infty$), and the Ricean distribution for $m = (K+1)^2/(2K+1)$ where $K$ is the Ricean factor. When the envelope of the received signal is Nakagami-$m$ distributed, the instantaneous SNR is gamma distributed with shape parameter $m$ and rate parameter $\snr^{-1}$, thus its Mellin transform is simply $\mathcal{M}_{g(\gamma)}(s) = \snr^{-m}\cdot U(m,s+m,\snr^{-1})$.

\subsubsection{MIMO eigen-beamforming}
 Consider now that the receiver is equipped with $N$ receive antennas and we perform eigen-beamforming at both transmitter and receiver ends. In order to maximize the SNR at the receiver, the transmit weighting vector $\mathbf{w}$ is selected to be the eigenvector of the Wishart matrix $\mathbf{H}^\herm\mathbf{H}$ which corresponds to the largest eigenvalue $\phi_{\rm max}$ of $\mathbf{H}^\herm\mathbf{H}$, i.e. $\gamma_{\rm mimo} = \snr\phi_{\rm max}$.
 
 \begin{cor}\label{cor:mimo_MRT}
 	The Mellin transform of the service process in the case of MIMO MRT in Rayleigh fading is given by
 	\begin{equation}
 	\setM_{g(\gamma)}^\MIMO(s) = R\sum_{i=1}^{N}\sum_{m=M-N}^{(M+N)i-2i^2}m!\cdot c_{i,m}\left(\frac{1}{\zeta}\right)^{m+1}\cdot U\left(m+1,m+1+s,i/\zeta\right)
 	\end{equation}
 	where $R =(\prod_{k=1}^{N}(N-k)!(M-k)!)^{-1}$ and coefficients $c_{i,m}$ can be obtained from \cite{Chen2005}; for the perfect CSI case, they have been tabulated for some values of $\{N,M\}$ \cite[Table I-IV]{Dighe2003}. This result is a direct consequence of \cite[Eq. 8]{Chen2005} and (\ref{eq:mellin_miso_tricomi}).
 \end{cor}
 
 Note that since the largest eigenvalue of the complex Wishart matrix (or equivalently the maximum singular value of $\mathbf{H}$) is bounded by $\frac{\|\mathbf{H}\|^2_{\rm F}}{\min(M,N)} \leq \phi_{\rm max} \leq \|\mathbf{H}\|^2_{\rm F}$
 and $\|\mathbf{H}\|^2_{\rm F} \sim \mathrm{Gamma}(MN,1)$, simple upper and lower bounds for the Mellin transform of MIMO eigen-beamforming can be obtained particularizing Theorem~\ref{Th1:main}.
\section{Delay with Large Blocklength: Asymptotic Analysis} \label{sec:perf_shannon_asym}
In the previous section, we have provided analytical expressions for the Mellin transform of the service process and the kernel for various multi-antenna diversity techniques. The exact results are mainly given in terms of special functions and alternating series. To explore further the delay performance of MISO MRT, we derive in this section simplified expressions for various asymptotic regimes: low/high SNR and large $M$. Additionally, we obtain a general result for a Gaussian distributed service process; as we will show, the MISO MRT service process converges to this distribution as $M$ grows large.
 
\subsection{High SNR regime} \label{sec:highPower}
We study here how latency constraints affect the MISO performance at high SNR. We assume this implies also large $\zeta$, which is true as long as $\sigma_\mathrm{e}^2$ does not increase\footnote{As a matter of fact, most frequently and in practice $\sigma_\mathrm{e}^2\propto 1/\snr$.} with the SNR (see (\ref{eq:snr_is_gamma})).
\begin{cor}
	In the high SNR regime, the Mellin transform of the service process scales as 
\begin{eqnarray}
\mathcal{M}_{g(\gamma)}^\HSNR(s) = \left\{
\begin{array}{ll}
	\zeta^{s-1}\frac{\Gamma(s+M-1)}{\Gamma(M)} & s > 1- M \\
	\zeta^{-M}\frac{\Gamma(1-s-M)}{\Gamma(1-s)} & s \leq 1- M \\
	\zeta^{-M}\frac{\log\zeta - \psi(M)}{\Gamma(M)} & s = 1-M \\
\end{array}
\right.
\end{eqnarray}
where $\psi(x)$ denotes the Digamma function \cite[Sec. 6.3]{Abramowitz1964}.
\end{cor}

\begin{IEEEproof}
The three branches are obtained after direct application of the asymptotic properties of the $U(a,b,z)$ function listed in \cite[Sec. 13.5]{Abramowitz1964}. The first branch can be also derived by considering the approximated service process $s_i\approx \log(\gamma_i)$, which gives that $\mathcal{M}_{g(\gamma)}^\HSNR(s) =  \zeta^{s-1}\Gamma(s+M-1)$ for $s+M > 1$.
\end{IEEEproof}

\subsection{Low SNR regime} \label{sec:lowPower}
At low SNR, the service process can be approximated as $s_i \approx \gamma_i$ and the following result is obtained.
\begin{cor}\label{cor:lsnr}
	In the low SNR regime, the Mellin transform of the service process is approximately given as  
	\begin{eqnarray}\label{eq:melling_lsnr}
	\mathcal{M}_{g(\gamma)}^\LSNR(s) = (1-(s-1)\zeta)^{-M}, \qquad s<\zeta^{-1}-1.
	\end{eqnarray}
\end{cor}
\begin{IEEEproof}
At low SNR, we use the first order Taylor series expansion $\log(1+x) \approx x$. In that case, the service process can be approximated as $s_i \approx \gamma_i$, which gives that $g(\gamma)\approx e^\gamma$, and in consequence
\begin{eqnarray}
\mathcal{M}_{g(\gamma)}^\LSNR(s) = \ex\left[{e^\gamma}^{(s-1)}\right] =  (1-(s-1)\zeta)^{-M}
\end{eqnarray}
using the moment generating function (MGF) of a gamma random variable.
\end{IEEEproof} 

\subsection{Large antenna regime} \label{sec:largeAnt}
The distribution of the mutual information of a Rayleigh fading MIMO system is generally rather complicated. For this reason, approximations have been used in the literature. For example, in the large antenna regime ($M \to \infty$) and using the Central Limit Theorem (CLT), it can be shown that the distribution of the mutual information $\mathcal{I}$ converges to a Gaussian distribution; see for instance \cite{Hochwald2004} and references therein. Using similar arguments here, we can obtain simpler expressions for the Mellin transform of the service process. In general, we can obtain results of the form
\begin{eqnarray}
M^{\alpha}\frac{\mathcal{I} - \mu}{\sigma_M} \stackrel{d}{\to} \mathcal{N}(0,1)
\end{eqnarray}
where convergence is in distribution, $\mu = \mathbb{E}(\mathcal{I})$, $\sigma_M$ is a variance term, and $\alpha$ is a measure of the convergence speed (normally 0.5). This means that, for large $M$, an accurate approximation of the distribution is given by $\mathcal{I} \sim \mathcal{N}(\mu,\sigma^2)$ with $\sigma^2 = \sigma_M^2/M^{\alpha}$. The mean and the variance terms can be obtained in closed form from \cite{McKay2008}. Note that, for brevity, throughout this section we will use the natural logarithm, and thus all rates are in nats.

Thanks to the CLT arguments and the Gaussian approximation of the service process, i.e. $s_i\approx\mathcal{I}_i$, we arrive at the following result.

\begin{thm}\label{th:gaussian}
	The Mellin transform of a service process with rate following a Gaussian distribution with mean $\mu$ and variance $\sigma^2$ is given by
	\begin{equation}\label{eq:mellin_gaussian}
	\mathcal{M}_{g(\gamma)}^{\as}(s) = e^{(s-1)\mu+(s-1)^2\frac{\sigma^2}{2}}.
	\end{equation}
\end{thm}
\begin{IEEEproof}
The service is given in terms of $\log(g(\gamma))$ (bit domain), thus we have that $g(\gamma) = e^{\mathcal{I}_i}$, and the result follows immediately by solving
\begin{eqnarray}
	\mathcal{M}_{g(\gamma)}^{\as}(s) &=& \ex\left[e^{(s-1)\mathcal{I}_i}\right] \\
	&=& \int_{-\infty}^{\infty}e^{(s-1)x}f_\mathcal{I}(x)\,\mathrm{d}x \\
	&=& e^{(s-1)\mu+(s-1)^2\frac{\sigma^2}{2}}.
\end{eqnarray} 
\end{IEEEproof}

\begin{thm}\label{th:hardening}
	For the MISO MRT case, as the number of antennas grows large, we have
	\begin{equation}
	\lim_{M\to\infty}\setM_{g(\gamma)}^{\as}(s) \to (1+\zeta M)^{s-1}.
	\end{equation}
\end{thm}
\begin{IEEEproof}
The mutual information can be written as $\mathcal{I} = \log(1+\gamma)$. Rewriting (\ref{eq:mellin_gaussian}) and applying Jensen's inequality
\begin{eqnarray}
\setM_{g(\gamma)}^{\as}(s) &\approx& e^{(s-1)\ex\left[\log(1+\gamma)\right]}\cdot e^{(s-1)^2\frac{\sigma2}{2}} \\
									&\leq&e^{(s-1)\log(1+\ex\left[\gamma\right])}\cdot e^{(s-1)^2\frac{\sigma2}{2}} \\
									&=& (1+\zeta M)^{s-1}\cdot e^{(s-1)^2\frac{\sigma2}{2}} \\
									&\stackrel{M\to\infty}{=}&(1+\zeta M)^{s-1}
\end{eqnarray}
where the last equality follows from the fact that $\lim\limits_{M\to\infty}\sigma^2 = 0$ \cite{Hochwald2004,McKay2008}.
\end{IEEEproof}
It can be shown that the bound is asymptotically tight (using Prohorov-Le Cam theorem, continuous mapping theorem, and Chebyshev inequality), but the proof is standard and is omitted for the sake of brevity. Furthermore, the asymptotic convergence can be obtained without resorting to the Gaussian approximation by showing that convergence in distribution implies convergence in $\mathcal{M}_{g(\gamma)}(s)$. Let $y_1, y_2, \ldots$ be a sequence of positive random variables that converges in distribution to a positive random variable $y$. For $s > 0$, we have $\displaystyle \lim_{M\to\infty}\mathcal{M}_{y_i}(s) = \mathcal{M}_{y}(s)$. By Lebesgue's dominated convergence theorem and $\frac{\left\|\mathbf{h}\right\|^2}{\mathbb{E}[\left\|\mathbf{h}\right\|^2]} \stackrel{\mathcal{P}}{\to} 1$, we have that $\displaystyle \lim_{M\to\infty}\mathcal{M}_{g(\gamma)}(s) = (1+\rho M)^{s-1}$.

Interestingly, we observe that for large $M$, $\setM_{g(\gamma)}^{\mathrm{as}}(s)\sim (\zeta M)^{s-1}$, which is related to the so-called channel hardening effect, i.e. the channel behaves equivalently to an AWGN channel with SNR $\zeta M$. In the low SNR regime, the number of transmit antennas affects linearly the service process, while at high SNR, the Mellin transform of the service process grows superlinearly with $M$ (for $s > 1$).

The approximation $s_i \sim \mathcal{N}(\mu,\sigma^2)$ allows us to simplify the delay violation probability expression (\ref{eq:pv_tricomi}), however its relevance and applicability goes beyond, as it allows analyzing the delay violation probability of any system whose service rate can be approximated by a Gaussian random variable. Additionally, it provides very simple expressions for the effective capacity, as we show next.

\subsection{Effective Capacity} \label{sec:effCapa}
Effective capacity is defined as the maximum constant arrival rate that a system can support given a QoS requirement $\theta$ \cite{WuNegi03_EC}. A byproduct of the delay analysis using MGF-based stochastic network calculus is that we can obtain expressions for the effective capacity $\setR$ by noticing that
\begin{equation}
\setR(\theta) \bydef -\frac{1}{\theta}\log\setM_{g(\gamma)}(1-\theta), \quad \theta>0.
\end{equation}
As an example, taking the normalized logarithm in (\ref{eq:mellin_miso_tricomi}), we can recover the effective capacity results in \cite{Matthaiou2012}. However, to assess the effect of multiple antennas in the delay-constrained performance, simpler expressions would be beneficial. We thus focus on the Gaussian approximation in Sec.~\ref{sec:largeAnt} and obtain
\begin{eqnarray}
\setR^{\mathrm{as}}(\theta) &=& \mu-\frac{\theta}{2}\sigma^2 \\
&=& \ex\left[\log(1+\gamma)\right]-\frac{\theta}{2}\var\left[\log(1+\gamma)\right].
\end{eqnarray}
As expected, the effective capacity converges to the ergodic capacity in the absence of delay constraints ($\theta=0$). For general $\theta$, there is a penalty on the achievable rate that is proportional to the variance of the instantaneous rate $\log(1+\gamma)$. This implies that, as the number of antennas tends to infinity, such penalty vanishes because the variance of the rate tends to zero \cite{Hochwald2004} and the effective capacity does not decay as $\theta$ increases.

\section{Delay Analysis with Finite Blocklength} \label{sec:perf_FB}
We investigate now the effect of finite blocklength in the service process and the delay performance. As explained in Sec.~\ref{sec:sys_model}, at finite blocklength there is always a probability of error $\epsilon$ and a rate loss as compared to Shannon capacity; in case of transmission errors, the offered service is zero. Therefore, the service process can be modeled as $s_i = R_i(n,\epsilon)\cdot Z_i$, where the coding rate $R_i(n,\epsilon)$ is approximated using (\ref{eqn:RateFB}) and $Z_i$ is a Bernoulli random variable, being one in case of successful transmission (with probability $1-\epsilon$), and zero otherwise; in this work, we assume independence between $Z_i$ and $\gamma_i$ (i.e. non-varying $\epsilon$ with SNR). 

Using (\ref{eqn:RateFB}) and lower bounding the achievable rate by zero (for very low SNR values), the Mellin transform of $g(\gamma_i,Z_i)$ is given by 
\begin{eqnarray}
\mathcal{M}_{g(\gamma_i,Z_i)}^\FB(s) = (1-\epsilon)\mathcal{M}_{q(\gamma_i)}(s) + \epsilon,
\end{eqnarray} 
where $q(\gamma_i) = \max\left(\frac{1+\gamma_i}{e^{\sqrt{V(\gamma_i)}F}},1\right)$ with $F = n^{-1/2}Q^{-1}(\epsilon)$, and $\mathcal{M}_{q(\gamma_i)}(s)$ is given by
\begin{eqnarray}
\mathcal{M}_{q(\gamma_i)}(s) & = & \ex\left[\max\left(\left(\frac{1+\gamma_i}{e^{\sqrt{V(\gamma_i)}F}},1\right)\right)^{s-1}\right] \nonumber \\
& = & \int_{0}^{\phi}f_{\gamma}(x){\rm d}x + \int_{\phi}^{\infty}\left(\frac{1+\gamma_i}{e^{\sqrt{V(\gamma)}F}}\right)^{s-1}f_{\gamma}(x){\rm d}x \nonumber \\
& = & \frac{\gamma(M,\phi/\snr)}{\Gamma(M)} + \mathcal{I}(c,s,F)
\end{eqnarray}
where $\phi = e^{F'}-1$ is the point where $\max{q(\gamma_i)} >1$.

The major difficulty in deriving $\mathcal{M}_{q(\gamma_i)}(s)$ is the fact that the channel dispersion $V(\gamma)$ depends on $\gamma_i$. In \cite{Al-Zubaidy15_FiniteBlock}, an infinite-order Taylor series expansion for $\sqrt{V(\gamma)}$ and the series expansion of the exponential function are used. These results can be easily extended in the MISO case, however the expression would be even more involved due to gamma distributed channel gains, providing little or no insight. For that, we numerically evaluate the integral when required and focus on a simpler asymptotic expression at high SNR.

\subsection{High SNR regime} \label{sec:highPowerFB}
At high SNR, the channel dispersion can be approximated as $V(\gamma) \approx 1$ and the Mellin transform of $q(\gamma_i)$ is approximately 
\begin{eqnarray}
\mathcal{M}_{q(\gamma_i)}(s) & \approx & \ex\left[\max\left(\left(\frac{1+\gamma_i}{e^{F}},1\right)\right)^{s-1}\right] \nonumber \\
& = & \int_{0}^{e^F-1}f_{\gamma}(x){\rm d}x + \int_{e^F-1}^{\infty}\left(\frac{1+\gamma_i}{e^{F}}\right)^{s-1}f_{\gamma}(x){\rm d}x \nonumber \\
& = & \frac{\gamma(M,(e^F-1)/\snr)}{\Gamma(M)} \nonumber \\ &+&\frac{e^{1/\zeta}\zeta^{-M}}{\Gamma(M)e^{F(s-1)}}\sum_{j=0}^{M-1}\binom{M-1}{j}(-1)^{M-1-j}\zeta^{j+s}\Gamma\left(j+s,e^F\zeta^{-1}\right)
\end{eqnarray}
where for the last equality we have followed the same procedure used to obtain Theorem~\ref{Th2:main} (see Appendix~\ref{sec:app_Th2main}). Note that for $F=0$, we recover the Mellin transform of the service process with infinite blocklength.
This approximation, which requires only widely used standard special functions, is easier to evaluate numerically. In Section~\ref{sec:num_results} we show that its accuracy is satisfactory even for moderate values of SNR.

\section{Numerical Results} \label{sec:num_results}
In this section, we provide numerical evaluation of the performance of MISO communication systems based on the above analysis. Unless otherwise stated, the duration of a slot is set to $T=1$~ms, the overhead is disregarded ($n_m \to 0$), and the blocklength is assumed to be $n=168$; consequently $B=n/\log 2\neq 1$, and we reincorporate this parameter into the equations.

\begin{figure*}
	\centering
	\includegraphics[scale=1.1]{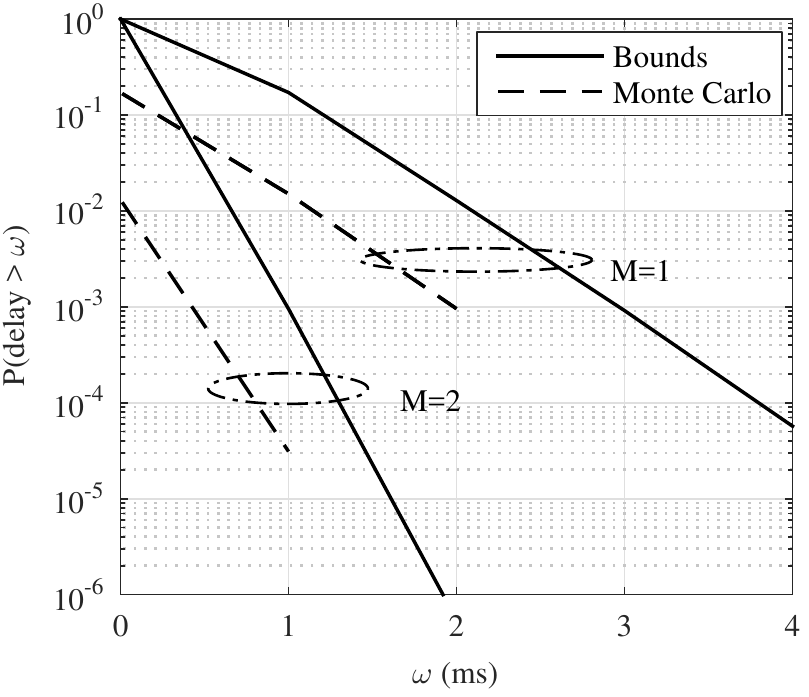}%
	\caption{Delay violation probability and associated bounds as a function of the target dealy, $\rho = 24$~kbps and $\snr=-2$~dB.}
	\label{fig:pv_mc}
\end{figure*}

We start by validating our analysis with Monte Carlo simulations. In Figure~\ref{fig:pv_mc}, we compare the delay violation probability and its bound with $\rho = 24$~kbps and $\snr = -2$~dB. We corroborate that the bounds follow the trend of the original curve, and we point out that the maximum difference in the x-axis seems to be of about $1$~ms.

\begin{figure*}
	\centering
	\includegraphics{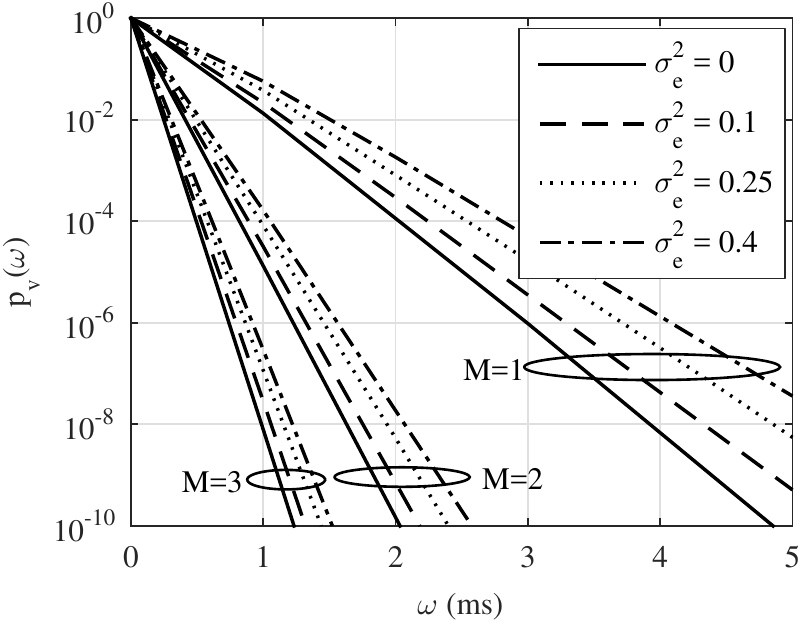}%
	\includegraphics{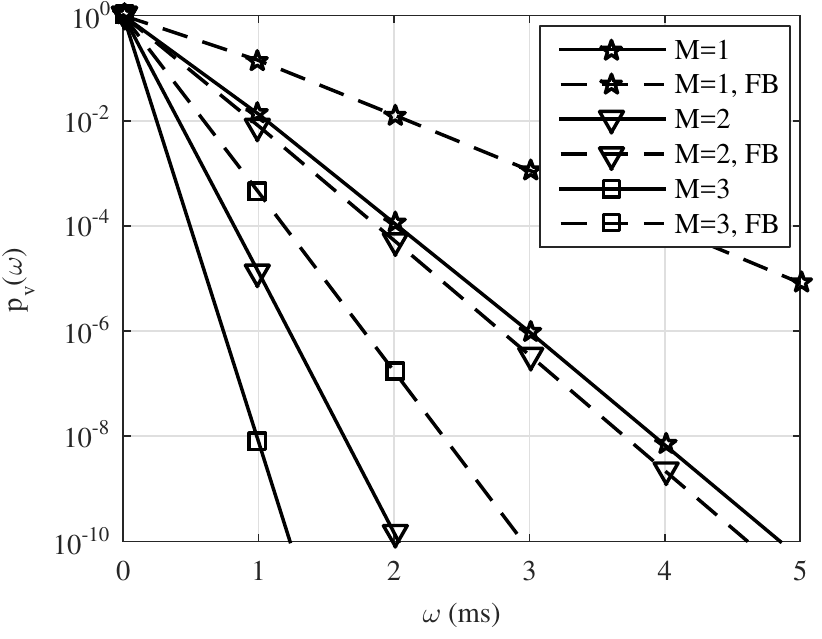}
	\caption{Delay violation probability bound as a function of the target delay, $\rho = 24$~kbps and $\snr=5$~dB. Curves labeled FB have been obtained using finite blocklength expressions.}
	\label{fig:pv_vs_w}
\end{figure*}

Figure~\ref{fig:pv_vs_w} plots the violation bound for MISO MRT as a function of the target delay $w$ with $\rho = 24$~kbps and $\snr = 5$~dB. The plot on the left shows the effect of varying the number of antennas and the accuracy of the CSI. We observe the strong decrease of the delay violation probability when increasing the number of antennas: with perfect CSI, the probability of exceeding 1~ms delay roughly decreases by three orders of magnitude when adding an extra antenna. On the other hand, the plot on the right depicts the difference between assuming finite and infinite blocklength; similar to \cite{Al-Zubaidy15_FiniteBlock}, we can see that such difference is remarkable, and that the Shannon model substantially overestimates the performance of the system.

\begin{figure*}
	\centering
	\includegraphics{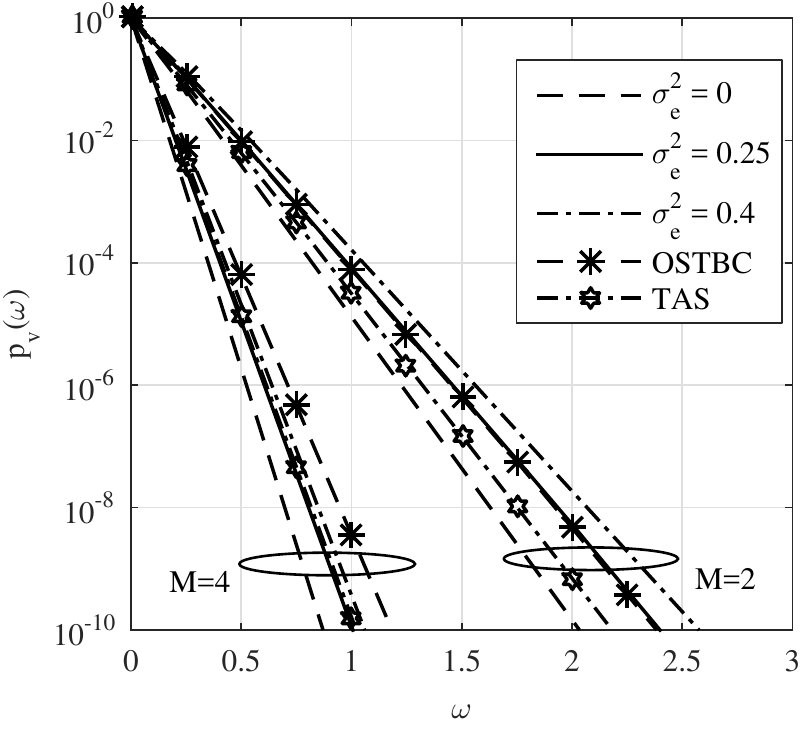}%
	\caption{Delay violation probability bound as a function of the target delay for different diversity techniques, $\rho = 24$~kbps and $\snr=5$~dB.}
	\label{fig:pv_div}
\end{figure*}

In Figure~\ref{fig:pv_div}, we compare the delay performance of MISO MRT with OSTBC and TAS. We can see that MRT generally performs better when the quality of the CSI is good: above a certain value of $\sigma_e^2$, TAS and OSTBC outperform MRT. The values at which this change takes place seem to be dependent on the number of antennas.

\begin{figure}
	\centering
	\includegraphics[scale=1.2]{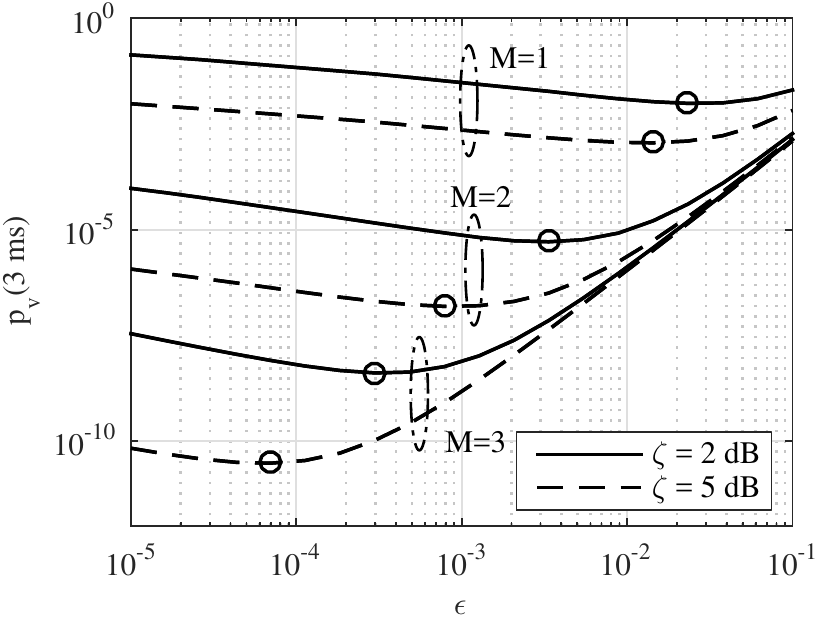}%
	\caption{Bound on the probability of exceeding $3$~ms of delay as a function of the block error rate $\epsilon$, finite blocklength analysis, $\rho = 24$~kbps. Circles mark the minimum of each curve.}
	\label{fig:pv_vs_epsilon}
\end{figure}

To obtain results with finite blocklength we must set an error probability $\epsilon$. In the experiment above, we have used $\epsilon = 10^{-2}$ for $M=1$, $\epsilon = 10^{-4}$ for $M=2$, and $\epsilon = 10^{-5}$ for $M=3$; these parameters have been set with the inspiration of Figure~\ref{fig:pv_vs_epsilon}, which illustrates the importance of choosing wisely $\epsilon$ depending not only on the SNR but also on the number of antennas.

\begin{figure}
	\centering
	\includegraphics{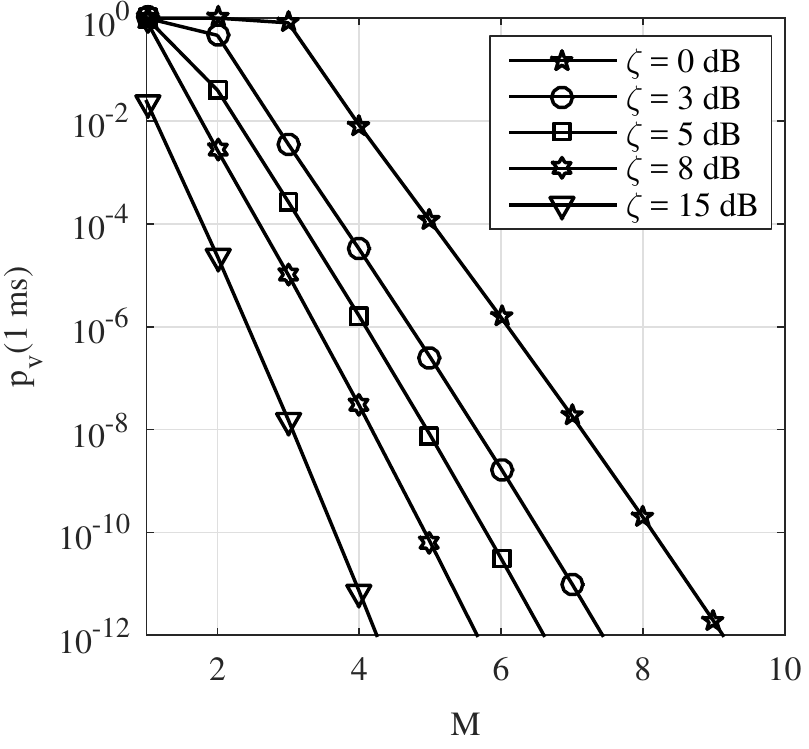}%
	\caption{Bound on the probability of exceeding $1$~ms delay as a function of the number of antennas, asymptotically large blocklength, $\rho = 256$~kbps.}
	\label{fig:pv_vs_M}
\end{figure}

In Figure~\ref{fig:pv_vs_M}, we investigate further the effect of adding antennas, and compare it to that of increasing the power. For a target delay of $1$~ms at $0$~dB, we can see that going from three to four antennas seems to have only slightly less impact than doubling the power; at $5$~dB, however, this is not the case anymore: $3$~dB of extra power decrease the violation probability by one order of magnitude, but adding one antenna decreases it by two orders of magnitude.  

\begin{figure*}
	\centering
	\includegraphics{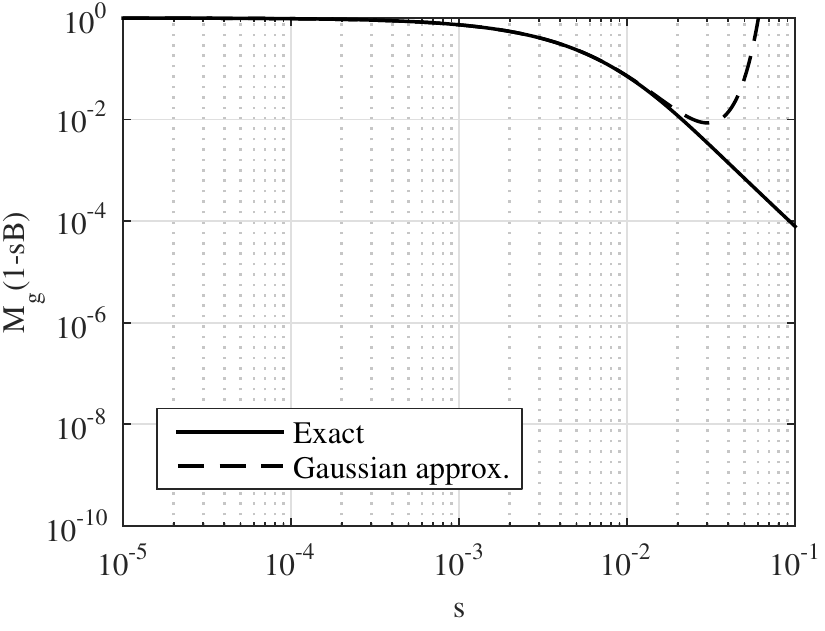}%
	\includegraphics{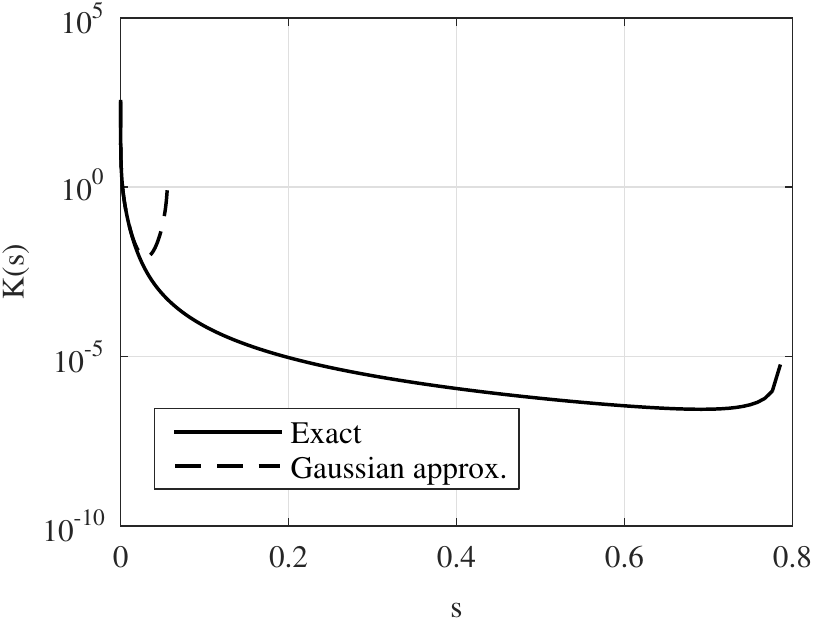}
	\caption{Mellin transform (left) and kernel (right) as a function of $s$, $M=3$, $\rho=20$~kbps, $\zeta = 0$~dB, $w=1$.}
	\label{fig:approx_M3}
\end{figure*}

\begin{figure*}
	\centering
	\includegraphics{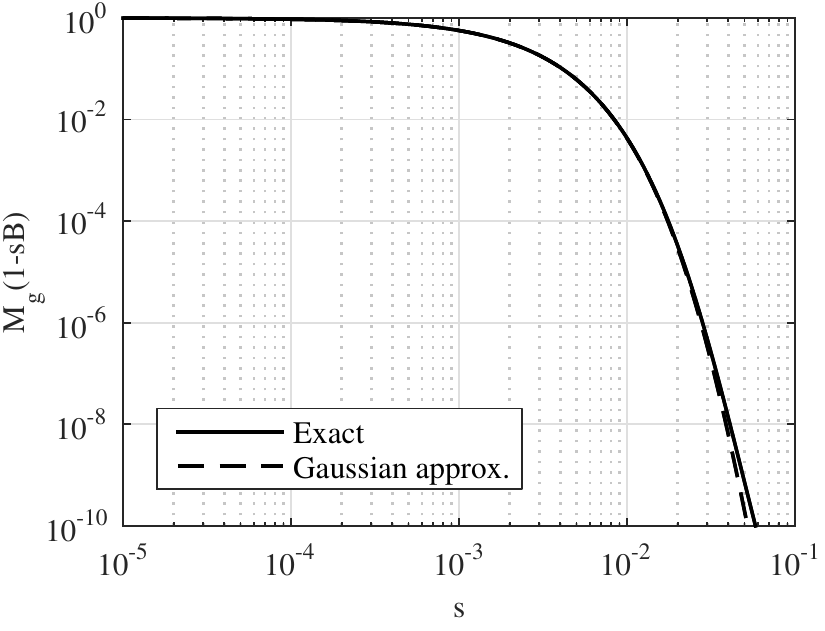}%
	\includegraphics{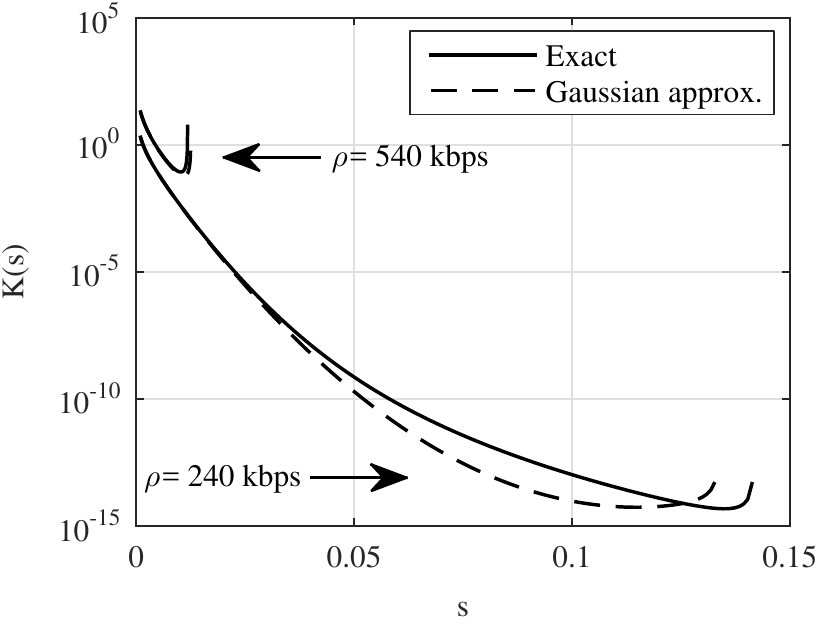}
	\caption{Mellin transform (left) and kernel (right) as a function of $s$, $M=10$, $\zeta = 0$~dB, $w=1$.}
	\label{fig:approx_M10}
\end{figure*}

As explained in Section~\ref{sec:perf_shannon}, it is important to have simple expressions of the kernel when possible. In Figure~\ref{fig:approx_M3} and Figure~\ref{fig:approx_M10}, we illustrate the accuracy of the Gaussian approximation for $M=3$ and $M=10$; as expected, the error is large for the former and negligible for the latter. This justifies the use of the much simpler expression (\ref{eq:mellin_gaussian}) whenever $M$ is relatively large.

\begin{figure*}
	\centering
	\includegraphics{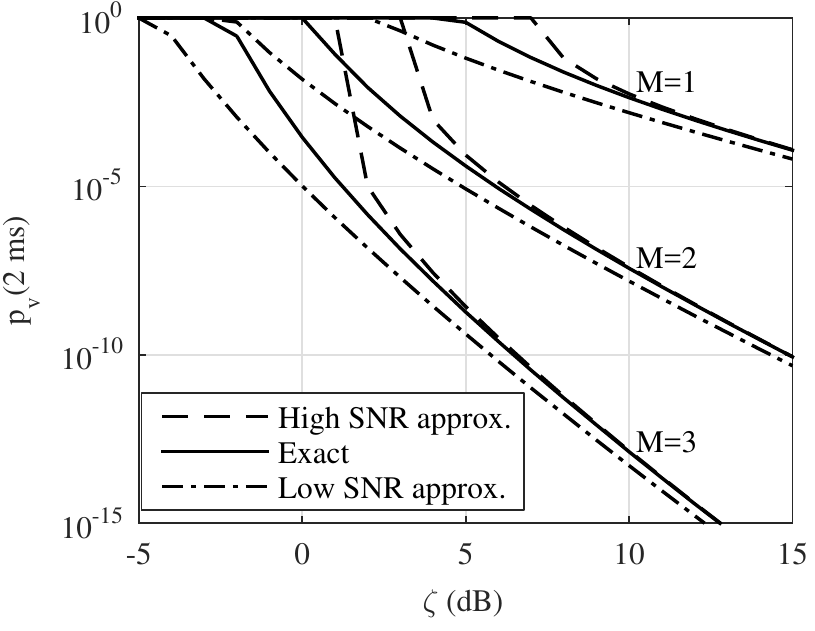}%
	\includegraphics{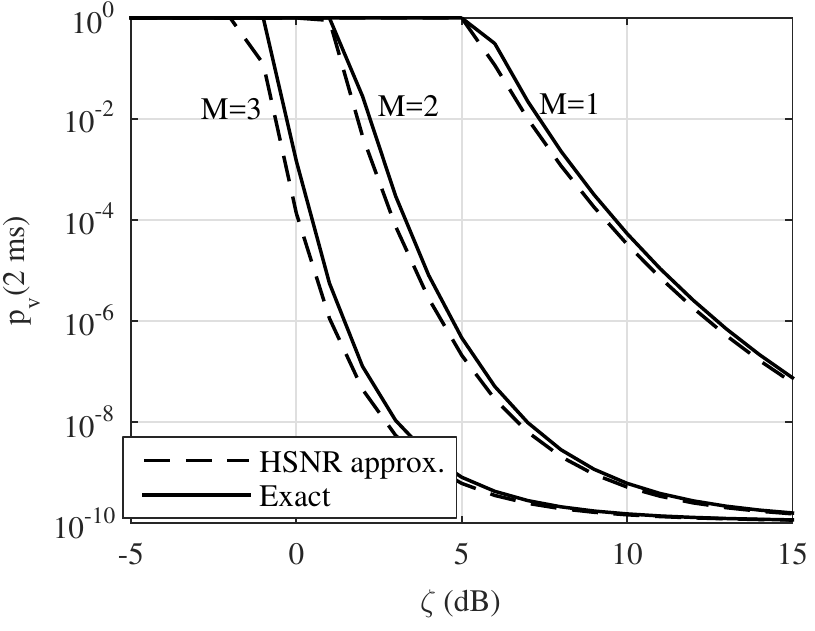}
	\caption{Bound on the probability of exceeding $2$~ms delay, asymptotically large blocklength (left) and finite blocklength (right) analysis, $\rho = 200$~kbps.}
	\label{fig:pv_vs_snr}
\end{figure*}

In Figure~\ref{fig:pv_vs_snr} (left) we test the accuracy of the high and low SNR approximations derived for the Shannon model in Section~\ref{sec:highPowerFB} and Section~\ref{sec:lowPower}. We can see that the high SNR approximation becomes asymptotically tight as the SNR increases, and that, remarkably, the low SNR approximation is reasonably accurate for most SNR values; this makes the low SNR approximation particularly interesting given its simplicity, see (\ref{eq:melling_lsnr}). Similarly, in Figure~\ref{fig:pv_vs_snr} (right) we show the accuracy of the high SNR approximation derived for the finite blocklength model in Section~\ref{sec:highPowerFB}.

\begin{figure}
	\centering
	\includegraphics[scale = 1.1]{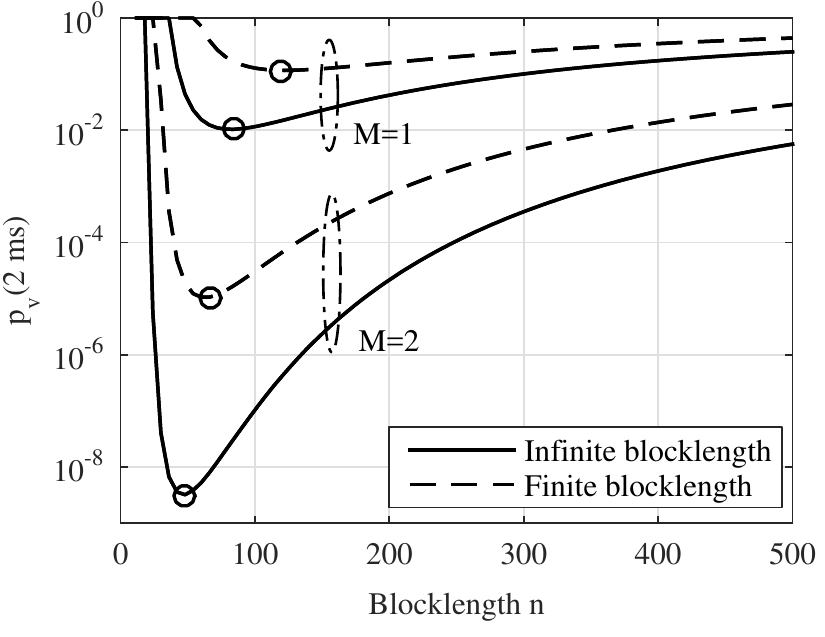}%
	\caption{Bound on the probability of exceeding $2$~ms delay as a function of the blocklength $n$, $\rho=150$~kbps, $\zeta = 10$~dB. Circles mark the minimum of each curve.}
	\label{fig:pv_vs_n}
\end{figure}

Finally, we show the effect of varying the blockelength $n$. We assume now a constant overhead of $n_m=64$~symbols, so that in each time slot a total of $n_m+n$ symbols are transmitted. The duration a time slot is now $(n+n_m)/168$~ms. As we can see from Figure~\ref{fig:pv_vs_n}, the delay performance heavily depends on the blocklength chosen, and the optimum value changes with the number of antennas.
\section{Conclusions} \label{sec:conclusions}
In this work, we characterized the delay performance of MISO diversity communications under statistical delay constraints. Using stochastic networks calculus, we derived a statistical characterization of the service process in multi-antenna fading channels and provided probabilistic delay bounds. We showed how the number of transmit antennas and transmit SNR may affect the delay performance. We also investigated the impact of imperfect CSI at the transmitter and finite blocklength channel coding on the delay performance of MISO transmit beamforming. MISO MRT is shown to reduce the delay violation probability as compared to single-antenna transmissions even with imperfect CSI. Nevertheless, as channel imperfections increase, other diversity-techniques, such as OSTBC and antenna selection, perform better than MRT in terms of delay violation probability. Future work could consider the effect of imperfect CSI at the receiver and limited feedback in FDD MIMO systems. Further extensions of this framework may include the analysis of MIMO spatial multiplexing, MIMO channels with co-channel interference, and multiuser MIMO systems.

\appendices

\section{Proof of Theorem~\ref{Th1:main}} \label{sec:app_Th1main}
Recall that $\gamma$ is gamma distributed with pdf $f_{\gamma}(x) = \frac{x^{M-1}e^{-\frac{x}{\zeta}}}{\Gamma(M)\zeta^M}, x \geq 0$. Then, we have that
\begin{eqnarray}
\mathcal{M}_{g(\gamma)}(s) &\bydef& \ex\left[(1+\gamma)^{s-1}\right] \\
						   &=& \int_0^\infty (1+x)^{s-1}f_{\gamma}(x)\,\mathrm{d}x \\
						   &=&\frac{1}{\zeta^M\Gamma(M)}\int_0^\infty (1+x)^{s-1}x^{M-1}e^{-x/\zeta}\,\mathrm{d}x \label{eq:tricomi_dev}\\
						   &\stackrel{(a)}{=}&\zeta^{-M}U(M,M+s,\zeta^{-1})
\end{eqnarray}
where (a) follows from the definition of Tricomi's confluent hypergeometric function \cite[Eq. 13.2.5]{Abramowitz1964} 
\begin{equation}
U(a,b,z) = \Gamma(a)^{-1}\int_0^\infty e^{-zt}t^{a-1}(t+1)^{b-a-1}\,\mathrm{d}t .
\end{equation} 
\section{Proof of Theorem~\ref{Th2:main}} \label{sec:app_Th2main}
Equation (\ref{eq:tricomi_dev}) can be rewritten as
\begin{eqnarray}
\mathcal{M}_{g(\gamma)}(s) &=&\frac{1}{\zeta^M\Gamma(M)}\int_0^\infty (1+x)^{s-1}x^{M-1}e^{-x/\zeta}\,\mathrm{d}x\\
&=& \frac{e^\frac{1}{\zeta}}{\zeta^M\Gamma(M)}\int_1^\infty (t-1)^{M-1}t^{s-1}e^{-t/\zeta}\,\mathrm{d}t
\end{eqnarray}
by applying change of variables $t = 1+x$. Now, since $M$ is a positive integer, we can use the binomial theorem to obtain
\begin{eqnarray}
\mathcal{M}_{g(\gamma)}(s)  &=& \frac{e^\frac{1}{\zeta}}{\zeta^M\Gamma(M)}\sum_{j=0}^{M-1}\binom{M-1}{j}(-1)^{M-1-j}\cdot\int_1^\infty t^{s+j-1}e^{-t/\zeta}\,\mathrm{d}t \\
&=& \frac{e^\frac{1}{\zeta}}{\zeta^M\Gamma(M)}\sum_{j=0}^{M-1}\binom{M-1}{j}(-1)^{M-1-j}\cdot\zeta^{j+s}\cdot\Gamma(j+s,\zeta^{-1}).
\end{eqnarray}
Or, alternatively,
\begin{equation}
\mathcal{M}_{g(\gamma)}(s) = e^\frac{1}{\zeta}\sum_{j=0}^{M-1}(-1)^{M-1-j}\cdot\zeta^{j+s-M}\frac{\Gamma(j+s,\zeta^{-1})}{\Gamma(M-j)\Gamma(j+1)}.
\end{equation}

\section{Proof of Theorem~\ref{Th3:main}} \label{sec:app_Th3main}
Suppose that $X_1, \ldots, X_n$ are $n$ independent continuous variates, each with cdf $F(x)$ and pdf $f(x)$. The pdf of the $r$-th order statistic $X_{(r)}$, $r = 1,\ldots,n$ is given by \cite{David_EVT}
\begin{eqnarray}
f_{(r)}(x) = \frac{1}{B(r,n-r+1)}F^{r-1}(x)[1-F(x)]^{n-r}f(x).
\end{eqnarray}
Therefore, the pdf of $\gamma_{\rm max} = \gamma_{(M)}$ is given by $f_{\gamma_{\rm max}}(x) = Mf_{\gamma}(x)F^{M-1}_{\gamma}(x)$. Since $\gamma \sim \mathrm{Exp}(1)$ in the case of TAS with pdf $f_{\gamma}(x) = e^{-x}$, we have that
\begin{eqnarray}
\mathcal{M}_{g(\gamma)}^\TAS(s) &\bydef& \ex\left[(1+\zeta\gamma_{\rm max})^{s-1}\right] \\
&=& \int_0^\infty (1+\zeta x)^{s-1}f_{\gamma_{\rm max}}(x)\mathrm{d}x \\
&=& M\int_0^\infty (1+\zeta x)^{s-1}f_{\gamma}(x)F^{M-1}_{\gamma}(x)\mathrm{d}x \\
&=& M\int_0^\infty (1+\zeta x)^{s-1}e^{-x}(1-e^{-x})^{M-1}\mathrm{d}x \\
&\stackrel{(a)}{=}& M\sum_{k=0}^{M-1}\binom{M-1}{k}(-1)^k\int_0^\infty (1+\zeta x)^{s-1}e^{-x(k+1)}\mathrm{d}x \\
&=& M\zeta^{s-1}\sum_{k=0}^{M-1}\binom{M-1}{k}(-1)^{k}\frac{e^{\frac{k+1}{\zeta}}}{(k+1)^s}\Gamma\left(s,\frac{k+1}{\zeta}\right)
\end{eqnarray}
where (a) follows from applying binomial theorem. 

\addcontentsline{toc}{chapter}{References}
\bibliographystyle{IEEEtran}
\bibliography{IEEEabrv,bib_snc,books_and_others}

\end{document}